\documentclass{article}
\usepackage{spconf,amsmath,graphicx}

\usepackage{spconf}
\usepackage{float}
\usepackage{times}
\usepackage{epsfig}

\usepackage{amsmath}
\usepackage{amssymb}
\usepackage{booktabs}
\usepackage[T1]{fontenc}
\usepackage{adjustbox}
\usepackage{lipsum}
\usepackage{amsmath}
\usepackage{algorithm}
\usepackage[noend]{algpseudocode}
\usepackage{tabularx,array,booktabs}

\usepackage{amssymb}
\usepackage{graphicx,epsfig,setspace,subfig,url,amsmath}
\usepackage{subfig}
\usepackage{longtable}
\usepackage{array}
\usepackage{amsmath}
\usepackage{mathtools}
\usepackage{caption}
\usepackage{multirow}
\usepackage[usenames, dvipsnames]{color}
\usepackage{soul}

 \normalsize

\usepackage{algorithm}
\usepackage[noend]{algpseudocode}
\usepackage{enumitem}
\setlist{nosep, leftmargin=14pt}

\title{Unsupervised Dual Adversarial Learning for Anomaly Detection in Colonoscopy Video Frames \sthanks{We acknowledge the support by the Australian Research Council through grant DP180103232 and the grant 2018/7063 - NALHN/THRF - Project Grant - 0006005804 and the TitanXp donated by NVidia. G.C. acknowledges the support from the Alexander von Humboldt-Stiftung for the renewed research stay sponsorship and Australian Heart Foundation (Ahrens Researcher Award). }}
\newcommand*\samethanks[1][\value{footnote}]{\footnotemark[#1]}
\name{\parbox{.8\linewidth}{\centering Yuyuan Liu$^\mathsection \sthanks{First two authors contributed equally to this work.}$   $\quad$ Yu Tian$^{\mathsection \|} \samethanks$  $\quad$ Gabriel Maicas$^\mathsection$ $\quad$ \newline Leonardo Zorron Cheng Tao Pu$^{\ddagger \delta}$ $\quad$  Rajvinder Singh$^\ddagger$ $\quad$ Johan W. Verjans$^{\mathsection\ddagger \|}$  $\quad$ Gustavo Carneiro$^\mathsection$}}
 \address {$^{\mathsection}$ Australian Institute for Machine Learning, School of Computer Science, University of Adelaide \\
 $^{\ddagger}$ Faculty of Health and Medical Sciences, University of Adelaide
 \\
  $^{\|}$ South Australian Health and Medical Research Institute \\
   $^{\delta}$ Department of Gastroenterology and Hepatology, Nagoya University }

\usepackage{changes}
	
\definecolor{yy_blue}{rgb}{45, 0, 179}

\begin{document}

\maketitle
\thispagestyle{empty}
\pagestyle{empty}

\begin{abstract}
\textit{The automatic detection of frames containing polyps from a colonoscopy video sequence is an important first step for a fully automated colonoscopy analysis tool.  Typically, such detection system is built using a large annotated data set of frames with and without polyps, which is expensive to be obtained.  In this paper, we introduce a new system that detects frames containing polyps as anomalies from a distribution of frames from exams that do not contain any polyps. 
The system is trained using a one-class training set consisting of colonoscopy frames without polyps -- such training set is considerably less expensive to obtain, compared to the 2-class data set mentioned above. 
During inference, the system is only able to reconstruct frames without polyps, and when it tries to reconstruct a frame with polyp, it automatically removes (i.e., photoshop) it from the frame -- the difference between the input and reconstructed frames is used to detect frames with polyps.
We name our proposed model as anomaly detection generative adversarial network (ADGAN), comprising a dual GAN with two generators and two discriminators.  
To test our framework, we use a new colonoscopy data set with 14317 images, split as a training set with 13350 images without polyps, and a testing set with 290 abnormal images containing polyps and 677 normal images without polyps. 
We show that our proposed approach achieves the state-of-the-art result on this data set, compared with recently proposed anomaly detection systems.}

\end{abstract}

\section{Introduction}
\label{sec:intro}

Colorectal cancer is considered to be one of the most harmful cancers -- current research suggests that it is the third largest cause of cancer deaths~\cite{siegel2014colorectal,tian2019one}. Early detection of colorectal cancer can be performed with the colonoscopy procedure for at-risk patients with symptoms like hemotochezia and anemia~\cite{bernal2017comparative}. Colonoscopy is based on the navigation of a small camera in the colon that enables doctors to classify and possibly remove or sample polyps, which are considered as the precursors of colon cancer~\cite{siegel2014colorectal}. The accurate detection of colon polyps may improve 5-year survival rate to over 90\%~\cite{siegel2014colorectal}.  Unfortunately, the accuracy of such manual detection varies substantially, leading to potentially missing detection that can have  harmful consequences for the patient~\cite{pu2018sa1908}. For instance, the false negative polyp detection can lead to a future colon cancer, which can be dangerous or even fatal~\cite{van2006polyp}. Therefore, automated detection of polyps is important in assisting doctors during a colonoscopy exam.

\begin{figure}[t!]
\small
\begin{center}
 \includegraphics[width = \linewidth]{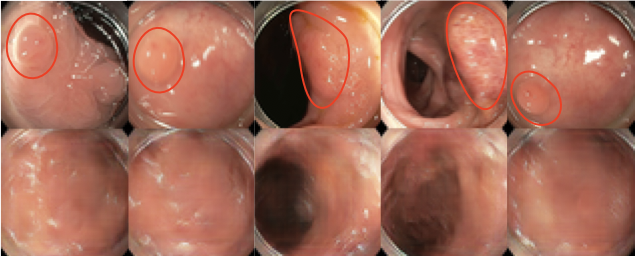}
\end{center}
\caption{Top row shows test images containing polyps (highlighted with a red ellipse), which are considered to be anomalies in our framework. Bottom row shows the reconstructed images by our ADGAN model, which deviate with their top row input images leading to high reconstruction errors. Note that given that the ADGAN model was trained with images without polyps, it is biased to reconstruct images without polyps, as clearly seen in these examples.
}
\label{fig:intro}
\vspace{-10pt}
\end{figure}
The automated polyp detection starts with the identification of frames containing polyps.  Typically, such detection system consists of a 2-class classifier, trained with images containing polyps and images that do not contain polyps.  The acquisition of such training set is expensive, requiring the manual annotation of large amounts of images for both classes.  Furthermore, given the intrinsic variations in the visualisation of polyps, it is challenging to collect a training set that is rich enough to thoroughly represent the class of images that contain polyps~\cite{tian2019one}.
To solve these two issues, this detection problem can be re-formulated as an anomaly detection problem, generally designed as a one-class classification problem that relies on a training set containing images that do not show the anomaly to be detected (i.e., the negative images)~\cite{camps2018one}. Such classification approach generally does not scale well with the size of the training set, reducing its applicability to large-scale medical image analysis problems. An alternative approach that addresses this scalability issue is based on a reconstruction model that first trains an encoder-decoder with negative images. Such model will produce high reconstruction errors for positive images (i.e., images containing polyps) during testing stage since only negative images were used during training~\cite{gong2019memorizing,makhzani2015adversarial,zong2018deep} -- see Fig.~\ref{fig:intro}. Note that the approaches above were developed for non-medical image analysis problems.
These encoder/decoder approaches suffer from two issues: 1) the reliance on mean square error (MSE) loss to compute the distance between the reconstructed images and its original image, can only preserve local visual information~\cite{goodfellow2014generative, makhzani2015adversarial}; and 2) the latent space learned with encoder/decoder approaches can accurately reconstruct abnormal images with similar appearance features to normal images, leading to relatively small deviation between the distributions of normal and abnormal data. For instance, the model that learns with colon wall images can also reconstruct the abnormal colon wall image containing small polyps~\cite{perera2019ocgan}. 

Anomaly detection can also be based on generative adversarial network (GAN)~\cite{goodfellow2014generative}, which addresses the local visual information issue mentioned above with an adversarial training of a generator that tries to fool a discriminator to be confused with the classification between real and synthetic images.  However, unlike the encoder-decoder model,  GAN~\cite{goodfellow2014generative} cannot reconstruct an image based on a compressed latent variable from a given input image, and suffers from unstable training.
Schlegl et al.~\cite{schlegl2019f} train an encoder to directly map an input image to GAN's latent space and tackle the unstable training issue by replacing the vanilla GAN~\cite{goodfellow2014generative} with Wasserstein GAN (WGAN)~\cite{arjovsky2017wasserstein}. 
Nevertheless, the framework proposed in~\cite{schlegl2019f} adopts a two-stage training strategy (encoder and WGAN are trained separately). 
Furthermore, the issues of the encoder/decoder model mentioned above are not addressed in~\cite{schlegl2019f}. 


In this paper, we propose a new WGAN-based~\cite{arjovsky2017wasserstein} anomaly detection model that comprises two generators and two discriminators -- this model is named anomaly detection GAN (ADGAN). Comparing with its competing approaches, our proposed model can produce an explicitly constrained latent space using latent generator and discriminator, and take advantage of GAN's generation ability to preserve both global and local information of input data by combining MSE and binary cross entropy (BCE) losses to improve the performance of anomaly detection. 
We show that our ADGAN is more effective (by relying on a one-stage end-to-end training) and more accurate (for anomaly classification) than previous methods~\cite{schlegl2019f,perera2019ocgan}. We demonstrate that our method can reconstruct the abnormal images with polyps into normal images by automatically removing (i.e., 'photoshopping') the polyps (Fig.~\ref{fig:intro}). 
These results are demonstrated on a new colonoscopy data set, containing 14317 high-quality colonoscopy images. The training set contains 13250 normal (healthy) images without polyps and we use 100 normal images as validation set. The testing set contains 290 abnormal images with polyps and 677 normal images without polyps (i.e., 30\% of the testing images are abnormal). 


\section{Related work}
\label{sec:Related work}

Anomalies are defined as data that do not conform to the general distribution of normal data. In medical image analysis, anomaly detection can be used, for example, in the detection of abnormal lesions in normal tissue~\cite{schlegl2019f,perera2019ocgan}. 
Anomaly detection models are generally based on one-class classifiers~\cite{camps2018one}, encoder-decoder models~\cite{perera2019ocgan,masci2011stacked} and GAN approaches~\cite{schlegl2019f,perera2019ocgan}. 
One-class classifiers are generally based on Gaussian processes, which do not scale well with training set size~\cite{camps2018one}. 
Encoder/decoder and GAN strategies assume that the abnormal data cannot be reconstructed correctly during testing stage given a model trained only with normal data. 
The typical encoder-decoder model for anomaly detection learns a deep auto-encoder~\cite{masci2011stacked} from the normal data during training stage, and during testing, this model is expected to produce larger reconstruction error for abnormal inputs~\cite{zhao2017spatio}, hopefully containing the lesions of interest. The main issue with the encoder-decoder method is that the trained model tends to accurately reconstruct abnormal samples during testing, leading to relatively small deviation between the distribution of normal and abnormal images. 

GAN-based models usually involve a conditional GAN approach, where the generated normal image is produced conditioned on another normal image.  For instance, Liu et al.~\cite{liu2018future} proposed a method based on future frame prediction in a video sequence using a GAN-based framework trained with normal data only, and tested to distinguish normal and abnormal events on surveillance data set. This approach is not applicable to medical image, and in particular colonoscopy data, because future frames tend to be not as predictable from past frames in a sequence.  OCGAN~\cite{perera2019ocgan} was proposed to distinguish abnormal data using a framework comprised of a de-noising auto-encoder network, latent discriminator, visual discriminator and a classifier. 
Nonetheless, the experiment results of this work indicate that the model perform unsatisfactorily on complex image data (e.g., surveillance and medical images)~\cite{perera2019ocgan}. 
Another GAN-based anomaly detection, Anogan~\cite{schlegl2017unsupervised}, trains a DCGAN~\cite{radford2015unsupervised} network on normal (healthy) retinal OCT images. During testing, a computationally expensive iterative back-propagation process is run to produce the closest image to the input image. 
Schlegl et al.~\cite{schlegl2019f} addressed this large computational run-time issue by training an encoder after training a WGAN~\cite{arjovsky2017wasserstein} to speed up the inference time. Replacing the iterative back-propagation from Anogan to an efficient encoding mechanism reduces inference running time, but introduces an ineffective two-stage training process. 
Another problem with the training above is that the encoder is under-constrained given that the MSE loss only recovers local visual information and misses global information. 

\begin{figure*}[t]
\small
\begin{center}
 \includegraphics[width=.95\textwidth]{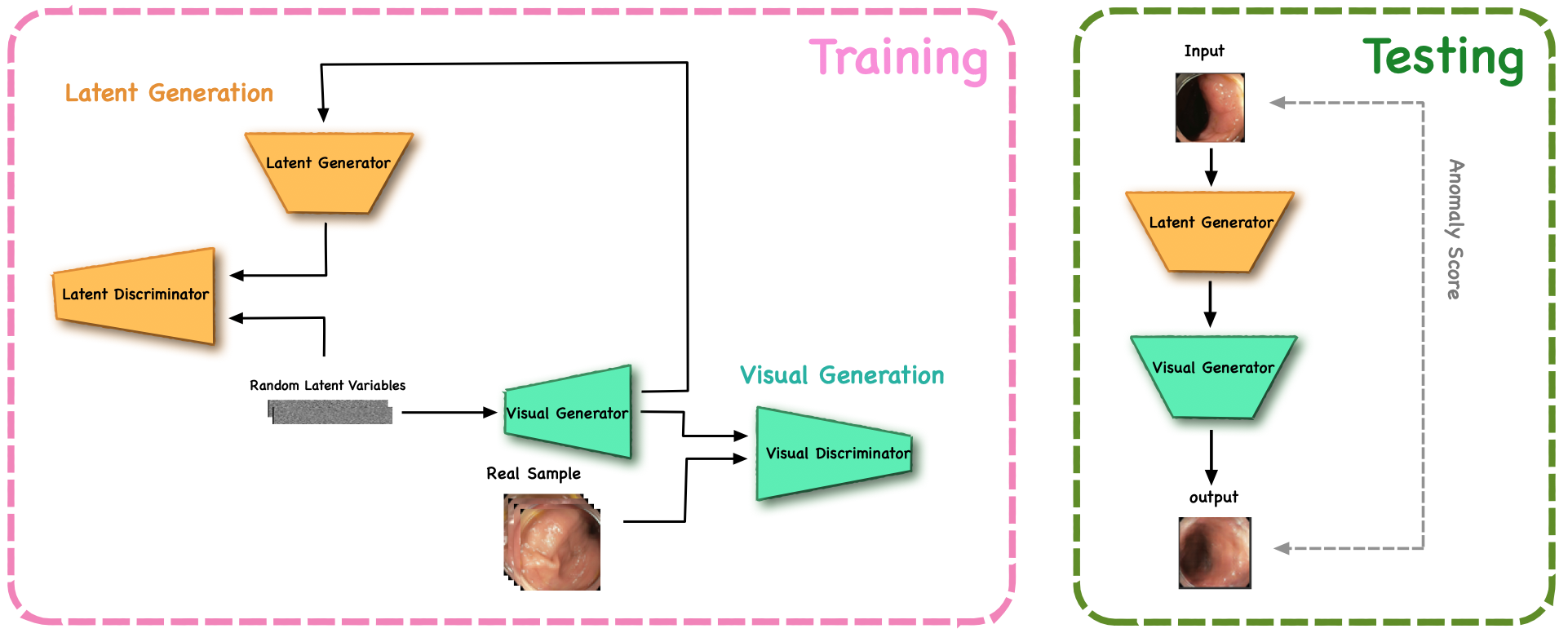}
\end{center}
\caption{Our proposed ADGAN model trains the visual generator, visual discriminator, latent generator and latent discriminator using adversarial training (left). During testing, the input image is processed by the latent generator and the produced latent embedding is used by the visual generator to produce the output image, which is then compared with the input image to compute the anomaly score.}
\label{fig:structure}
\end{figure*}

By taking the motivation from~\cite{schlegl2019f}, our proposed GAN framework, based on the Wasserstein GAN (WGAN)~\cite{arjovsky2017wasserstein},  resolves the issues mentioned above by an end-to-end (i.e., one-step) training of a dual GAN that uses a new loss function that minimises both the local and global reconstruction errors.
  
\section{Data Set and Methods}
\label{sec:method}
\subsection{Data Set} 
\label{sec:method_data}

The data set is obtained from 18 colonoscopy videos from 15 patients.  Video frames containing blurred visual information are removed using the variance of Laplacian method~\cite{he2006laplacian}. We then sub-sample consecutive frames by taking one frame every five frames because 
consecutive frames generally contain similar visual information that makes GAN training ineffective.  We also remove frames containing feces and water to improve the training efficiency (we plan to deal with such distractors in future work). As a result, the frames used for training and testing are sharp, clean (of feces and water), and discontinuous (in time domain).  

This data set is defined by $\mathcal{D} = \{ \mathbf{x}_i,d_i,y_i \}_{i=1}^{|\mathcal{D}|}$, where $\mathbf{x}:\Omega \rightarrow \mathbb R^3$ denotes a colonoscopy frame ($\Omega$ represents the frame lattice), $d_i \in \mathbb N$ represents patient identification\footnote{Note that the data set has been de-identified -- $d_i$ is useful only for splitting $\mathcal{D}$ into training, testing and validation sets in a patient-wise manner.}, $y_i \in \mathcal{Y} = \{ Normal, Abnormal \}$ denotes the normal healthy colorectal frames and abnormal colorectal frames that contains polyp.  
The distribution of this data set is as follows: 1) Training set: 13250 normal (healthy) images without any polyps; 2) Validation set: 100 normal (healthy) images for model selection; and 3) Testing set: 967 images, with 290 (30\% of the set) abnormal images with polyps and 677 (70\% of the set) normal (healthy) images without polyps.  Note that the patients in testing set do not appear in the training/validation sets and vice versa. This abnormality proportion (on the testing set) is commonly defined in other anomaly detection literature~\cite{perera2019ocgan}~\cite{schlegl2019f}.
These frames were obtained with the Olympus~\textregistered 190 dual focus colonoscopy. 


\subsection{Methods}
\label{sec:method_method}

Our proposed \textbf{ADGAN} is shown in Figure~\ref{fig:structure}, and comprises a visual generator, a visual discriminator, a latent generator and a latent discriminator.
Defining $\mathbf{z} \sim \mathbb{U}(-1, 1)$, with $\mathbf{z} \in \mathbb R^Z$ as the latent variable, the visual generator is defined by
\begin{equation}
    \hat{\mathbf{x}} = G_v( \mathbf{z}; \: \theta_v),
    \label{eq:G_v}
\end{equation}
where $\theta_v$ denotes the parameter vector of the generator, and $\hat{\mathbf{x}}:\Omega \rightarrow \mathbb R^3$ denotes the generated image.  Similarly, the latent generator is defined by
\begin{equation}
    \hat{\mathbf{z}} = G_l( \hat{\mathbf{x}}; \: \theta_l),
    \label{eq:G_l}
\end{equation}
where $\theta_v$ is again the parameter vector. During training, $G_v(.)$ from~\eqref{eq:G_v} generates fake images $\hat{\mathbf{x}}$ given $\mathbf{z}$ to fool the visual discriminator, defined as:
\begin{equation}
    r = D_v( \hat{\mathbf{x}}; \gamma_v),
    \label{eq:D_v}
\end{equation}
where $\gamma_v$ represents the discriminator parameters.  The generated image $\hat{\mathbf{x}}$ is then fed to the latent generator in~\eqref{eq:G_l} to produce a fake latent vector $\hat{\mathbf{z}}$ to fool the latent discriminator, defined as  
\begin{equation}
    r = D_l( \hat{\mathbf{z}}; \gamma_l),
    \label{eq:D_l}
\end{equation}
where $\gamma_l$ is the discriminator parameters.
The training process follows~\cite{gulrajani2017improved}, where we minimise the visual generation loss,
\begin{equation}
    \begin{aligned}
    l_{D_v} & \: = \: D_{v}(\mathbf{x}) - D_{v}(\hat{\mathbf{x}}) + \lambda(||\nabla D_v(\hat{\mathbf{x}})||_2 - 1)^2\\
    l_{G_v} & \: = \: -D_{v}(G_{v}(\mathbf{z}));
    \end{aligned}
\end{equation}
and the latent generation loss:
\begin{equation}
    \begin{aligned}
    l_{D_l} & \: = \: \log(D_{l}(\mathbf{z})) + \log(1-D_{l}(\hat{\mathbf{z}}))\\
    l_{G_l} & \: = \: \alpha \: \log(1-D_{l}(G_{l}(\hat{\mathbf{x}}))).
    \end{aligned}
\end{equation}
To generate realistic images, we also minimise the mean squared error (MSE) loss between the input and generated latent vectors, as in  $l_{MSE}=\beta \: ||\mathbf{z}-G_l(G_v(\mathbf{z}))||_2^2$. 
For training, the hyper-parameters $\alpha, \beta$ are estimated from the validation set within the range $[0.1, 10]$.
The training process consists of $N$ iterations, where the visual generator is trained for $T<N$ iterations, and then the whole model is trained for $(N - T)$ iterations.

During testing, given a sample $\mathbf{x}$, a latent vector $\hat{\mathbf{z}}$ is produced with~\eqref{eq:G_l}, which is then fed to the visual generator in~\eqref{eq:G_v} and the anomaly score is computed with:
\begin{equation}
    A(\mathbf{x}) = \Vert \mathbf{x} - G_{v}(G_{l}(\mathbf{x}) )  \Vert ^2_2. 
    \label{eq:A}
\end{equation}
Small anomaly score indicates normal samples and high anomaly score generally indicates abnormal samples, suggesting the presence of a polyp (see Fig.~\ref{fig:intro} for a few reconstructions examples produced by ADGAN). 

\section{Experiment}
\label{sec:Experiments} 

In this section, we validate our proposed ADGAN model using the data set described in Sec.~\ref{sec:method_data}. We compare our performance with other baseline approaches and state-of-the-art methods. We show that our model achieves state-of-the-art area under the ROC curve (AUC) results. 

\subsection{Experimental Setup}

We pre-process the original colonoscopy image from 1072 $\times$ 1072 $\times$ 3 resolution to 64 $\times$ 64 $\times$ 3 to reduce the computational cost of the training and inference processes. The model selection is done with  the validation set mentioned in Sec.~\ref{sec:method_data}.  This method is implemented using Pytorch~\cite{paszke2017automatic} and the code will be publicly available upon acceptance of the paper. We use Adam~\cite{kingma2015adam} optimiser during training with a learning rate of 0.0001. Our model has a similar backbone architecture as the other competing methods in 
Tab.~\ref{tab:result}. 
In particular, the visual generator and discriminator are based on the improved GAN~\cite{salimans2016improved} and use four residual convolution and four residual de-convolution layers, respectively~\cite{salimans2016improved}. 
The latent generator and discriminator are based on DCGAN~\cite{radford2015unsupervised} with three convolution/de-convolution layers. 
The number of filters per layer for our visual discriminator are (64, 128, 256, 512) (reverse order for visual generator). The number of filters per layer for our latent generator are (64, 128, 256, 512), and we use (256, 128, 64) as the number of filters per layer for our latent discriminator.
To train the model, we first train the  visual generator and discriminator for 80000 iterations while fixing the parameters of latent generator and discriminator. We then jointly train the whole framework for 20000 iterations, with a batch size of 64. 

\subsection{Anomaly Detection Results}

\begin{table}[ht!]
\small
\centering
\scalebox{1.1}{
\begin{tabular}{c|c}
\toprule
Methods        & AUC \\ \hline
DAE~\cite{masci2011stacked} & 0.6294\\
VAE~\cite{doersch2016tutorial} & 0.6478  \\
\hline
OC-GAN~\cite{perera2019ocgan}         & 0.5916   \\
f-AnoGAN(ziz)~\cite{schlegl2019f}  & 0.6376  \\
f-AnoGAN(izi)~\cite{schlegl2019f}  & 0.6638   \\
f-AnoGAN(izif)~\cite{schlegl2019f} & 0.6913   \\
\textbf{ADGAN}         & \textbf{0.7296}   \\ \bottomrule
\end{tabular}}
\caption{Comparison between our proposed ADGAN and other state of the art methods.}
\label{tab:result}
\end{table}

We compare the proposed ADGAN with state-of-the-art approaches, including OCGAN~\cite{perera2019ocgan}, f-anogan and its variants~\cite{schlegl2019f} that involve image-to-image MSE loss (izi), Z-to-Z MSE loss (ziz) and its hybrid version (izif). 
We also compare our method with some baseline approaches, including deep auto-encoder~\cite{masci2011stacked} and variational auto-encoder~\cite{doersch2016tutorial}. The anomaly score $A(\mathbf{x})$ in~\eqref{eq:A} is used to indicate the presence of polyps. For the encoder-decoder architecture comparison, the models adopt similar structure as our latent generator and latent discriminator. For GAN-based methods comparison, we use the same structure as our visual generator and latent discriminator with similar model capacity. We use area under the ROC curve (AUC) as the measurement for performance validation~\cite{schlegl2019f,perera2019ocgan}. As shown in Table~\ref{tab:result}, our ADGAN model outperforms other methods. 

\subsection{Image Reconstruction from ADGAN}

Figure~\ref{fig:intro} demonstrates the input images (first row) from testing set and their reconstructed images (second row) with our proposed ADGAN model. We manually mark the abnormal polyp lesions using red circles. Our model reconstructs the abnormal input images to their healthy versions, leading to substantial reconstruction error (anomaly score) due to visual differences. The normal images from testing set are generally reconstructed well producing small reconstruction errors (anomaly score).  

\section{Conclusions}
\label{sec:Conclusions} 

In conclusion, we proposed a GAN-based framework (ADGAN) for anomaly detection using one-class learning on a colonoscopy data set. The model was trained end-to-end and experiments show that our model achieved the state-of-the-art anomaly detection result. We solve the issues of mapping between input image and GAN's latent space using a second GAN model,
and proposed a new loss function that combines MSE loss, Wasserstein loss and standard BCE loss. In the future, we plan to extend our model to work with colonoscopy images showing feces and water, as explained in Sec.~\ref{sec:method_data}.
\bibliographystyle{IEEEbib}
\bibliography{refs}

\end{document}